\documentclass[5p, twocolumn, times]{elsarticle}

\usepackage{graphicx}
\usepackage{amsmath}   

\begin{document}

\begin{frontmatter}

\title{Beam Performance of Tracking Detectors with Industrially Produced GEM Foils}

\author[add1,add2]{F.~Simon\corref{cor}}
\ead{frank.simon@universe-cluster.de}
\author[add3]{J.~Kelsey}
\author[add6]{M.~Kohl}
\author[add4]{R. Majka}
\author[add3]{M.~Plesko}
\author[add3]{T.~Sakuma}
\author[add4]{N.~Smirnov}
\author[add5]{H.~Spinka}
\author[add3]{B.~Surrow}
\author[add5]{ D.~Underwood}

\cortext[cor]{Corresponding author}

\address[add1]{Max-Planck-Institut f\"ur Physik, M\"unchen, Germany}
\address[add2]{Excellence Cluster Universe, Technische Universit\"at  M\"unchen, Garching, Germany}
\address[add3]{Laboratory for Nuclear Science, Massachusetts Institute of Technology, Cambridge, MA, USA}
\address[add6]{Hampton University, Hampton, VA, USA}
\address[add4]{Physics Department, Yale University, New Haven, CT, USA}
\address[add5]{Argonne National Laboratory, Argonne, IL, USA}

\begin{abstract}
Three Gas-Electron-Multiplier tracking detectors with an active area of 10 cm $\times$ 10 cm and a two-dimensional, laser-etched orthogonal strip readout have been tested extensively in particle beams at the Meson Test Beam Facility at Fermilab. These detectors used GEM foils produced by Tech-Etch, Inc. They showed an efficiency in excess of 95\% and spatial resolution better than 70 $\mu$m. The influence of the angle of incidence of particles on efficiency and spatial resolution was studied in detail.

\end{abstract}


\end{frontmatter}

\section{Introduction}

Micro-pattern gas detectors have proven to be versatile devices for high resolution particle tracking. One of the most successful micro-pattern technologies is the Gas Electron Multiplier (GEM), introduced in 1996 at CERN \cite{Sauli:1997qp}. The GEM is a thin metal-clad insulator foil that is chemically perforated with a large number of small holes. Voltage applied across the foil generates strong electric fields in the holes, which leads to avalanche multiplication of electrons. Since the electron amplification occurs in the holes of the GEM foil and is separated from charge collection structures, the choice of readout geometries for detectors based on the GEM is very flexible. For tracking applications several GEM foils are cascaded to reach higher gain and high operating stability. Spatial resolutions of better than 70 $\mu$m have been demonstrated with triple GEM detectors \cite{Altunbas:2002ds}, with a material budget of significantly less than 1\% of a radiation length ($X_0$) per tracking layer (providing a 2D space point). 

To meet the increasing demand for GEM foils for research applications the establishment of commercial producers is desirable. A collaboration with Tech-Etch, Inc., based on an approved SBIR\footnote{Small Business Innovative Research, US-DOE funded program to foster collaboration of small companies and research institutions} proposal, has been formulated to provide a commercial source for GEM foils and to study the production of large area foils. GEMs produced by Tech-Etch, Inc. have been evaluated in detail both with optical methods and in test detectors to study geometrical uniformity as well as gain performance \cite{Simon:2007sk}. 

One application for the GEM foils produced by TechEtch, Inc. is the Forward GEM Tracker (FGT) \cite{Simon:2007fz} of the STAR Experiment  \cite{Ackermann:2002ad} at the Relativistic Heavy Ion Collider (RHIC). This approved upgrade will provide high-precision tracking at forward rapidity, covering the acceptance of the STAR endcap electromagnetic calorimeter (EEMC) \cite{Allgower:2002zy}. With this upgrade, the charge sign of high transverse momentum electrons and positrons from $W$ decays can be identified, which is crucial for the study of flavor-separated polarized quark distributions in the proton. To achieve this, a multi-layer low mass tracker with \mbox{$\sim$80 $\mu$m} spatial resolution or better is needed, making GEM detectors a good choice for this upgrade.

\section{Beam Test Setup}

In order to evaluate the performance of GEM foils produced by Tech-Etch in an application environment, a test detector based on the geometry used in the COMPASS experiment \cite{Altunbas:2002ds} has been developed. The detector was a triple GEM design with a two-dimensional projective strip readout. The active area, given by the size of the GEM foils, was 10 cm $\times$ 10 cm. The test detectors were designed to allow for easy replacement of individual foils. They had an aluminum bottom support plate with a machined-out section covering the active area of the detector, and plexiglass sides and top cover. Over the active area, the polyimide side of the cathode foil, a 50 $\mu$m kapton foil with 5 $\mu$m copper coating, was exposed. The detectors had a material budget of around 4\% $X_0$ each, mostly due to the bottom support plate made out of aluminum and due to the 2D readout board manufactured on a standard printed circuit board. A pre-mixed gas of Ar:CO$_2$ (70:30) was used for all measurements. 

The foils were powered from a single high voltage source through a resistor chain with equal voltage sharing between the three foils. The resistors were 2 M$\Omega$ over the drift, transfer and induction gaps and 1.2 M$\Omega$ over the GEM foils. The top of each GEM foil and the cathode were connected to the voltage divider via a 10 M$\Omega$ protection resistor to limit the current in the event of a discharge. The drift gap of the detector between the cathode foil and the top GEM was 3.2 mm, the transfer gap between the other foils and between the bottom GEM and the readout board were 2.2 mm. The nominal operation voltage of the detectors ranged from 3750 to 3800 V, corresponding to $\sim$ 385 -- 395 V across each GEM foil. This resulted in a total effective detector gain of approximately  \mbox{3\,500}. The gain of the detectors during data taking was chosen according to the dynamic range of the electronics used in the beam test, and was far below the maximum gain for stable operation  \cite{Simon:2007sk}.

\begin{figure}
\centering
\includegraphics[width=0.48\textwidth]{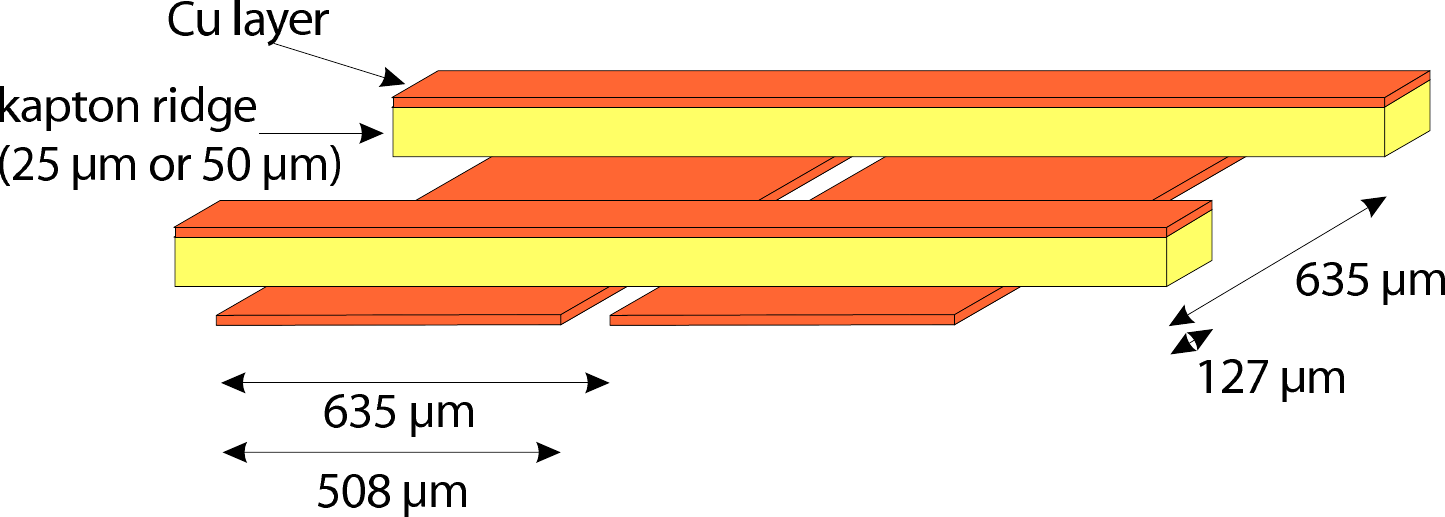}
\caption{The schematic structure of the laser-etched 2D orthogonal strip readout board. Two different versions of this board were used, one with 25 $\mu$m and one with 50 $\mu$m thick kapton ridges that define the vertical distance of the two strip layers. In the test beam setup, the horizontal (X) coordinate was read out by the bottom strips, the vertical (Y) coordinate was read out by the top strips for all three detectors.}
\label{fig:ReadoutBoard}
\end{figure}

Charge collection was done with a two-dimensional orthogonal strip readout, manufactured using laser etching techniques by Compunetics, Inc. . This readout structure, illustrated in Figure \ref{fig:ReadoutBoard}, had a strip pitch of 635 $\mu$m. The top strips were 127 $\mu$m wide and the bottom strips were 508 $\mu$m wide. Two vertical separations of the strip layers were used, one with 25 $\mu$m and one with 50 $\mu$m. This allowed the investigation of the effect of the board geometry on the charge sharing between coordinates. The strip readout has been fully integrated in a multi-layer printed circuit board that contained the signal distribution lines to the connectors for the front-end chips and the control lines from the front-end chip connectors to the connectors for the control units as well as the high voltage divider system.  For all detectors the horizontal (X) coordinate was read out by the bottom strips of the readout structure, while the vertical (Y) coordinate was read out by the top strips.

Before installation in the detectors all GEM foils were tested for electrical stability and underwent an optical analysis to establish their geometric parametes, using an automated high resolution scanning setup \cite{Becker:2006yc,Simon:2007sk}. Each of the  triple GEM test detectors was also evaluated with a $^{55}$Fe source to study gain uniformity and charging behavior \cite{Simon:2007sk}.

The test detectors were equipped with 6 APV25-S1 front-end chips each, with 64 channels per chip connected to readout strips. The strips were directly connected to the chip without an additional diode and resistor protection, as used for example in the COMPASS detectors \cite{Altunbas:2002ds}. The APV25-S1 chips were controlled and read out through a custom made module based on FPGAs directly on the detector. The off-detector data acquisition system was set up to mimic the STAR DAQ system \cite{Landgraf:2002zw}, using a STAR trigger/clock module and a modified DAQ unit from the STAR time of flight system. This unit received the data from the detector and communicated with a DAQ computer via a fiber link developed for the ALICE experiment at CERN. The APV chips were sampling the signal at a rate of 40 MHz and were read out in peak mode, meaning that a single time slice per event was recorded for each strip. The timing of the chips was set up so that this recorded time slice corresponded to the maximum signal amplitude. The dynamic range of the readout electronics for physics signals with the configuration used for the beam tests was approximately 600 ADC channels, corresponding to $\sim \, 10^5$ electrons.

\begin{figure}
\centering
\includegraphics[width=0.48\textwidth]{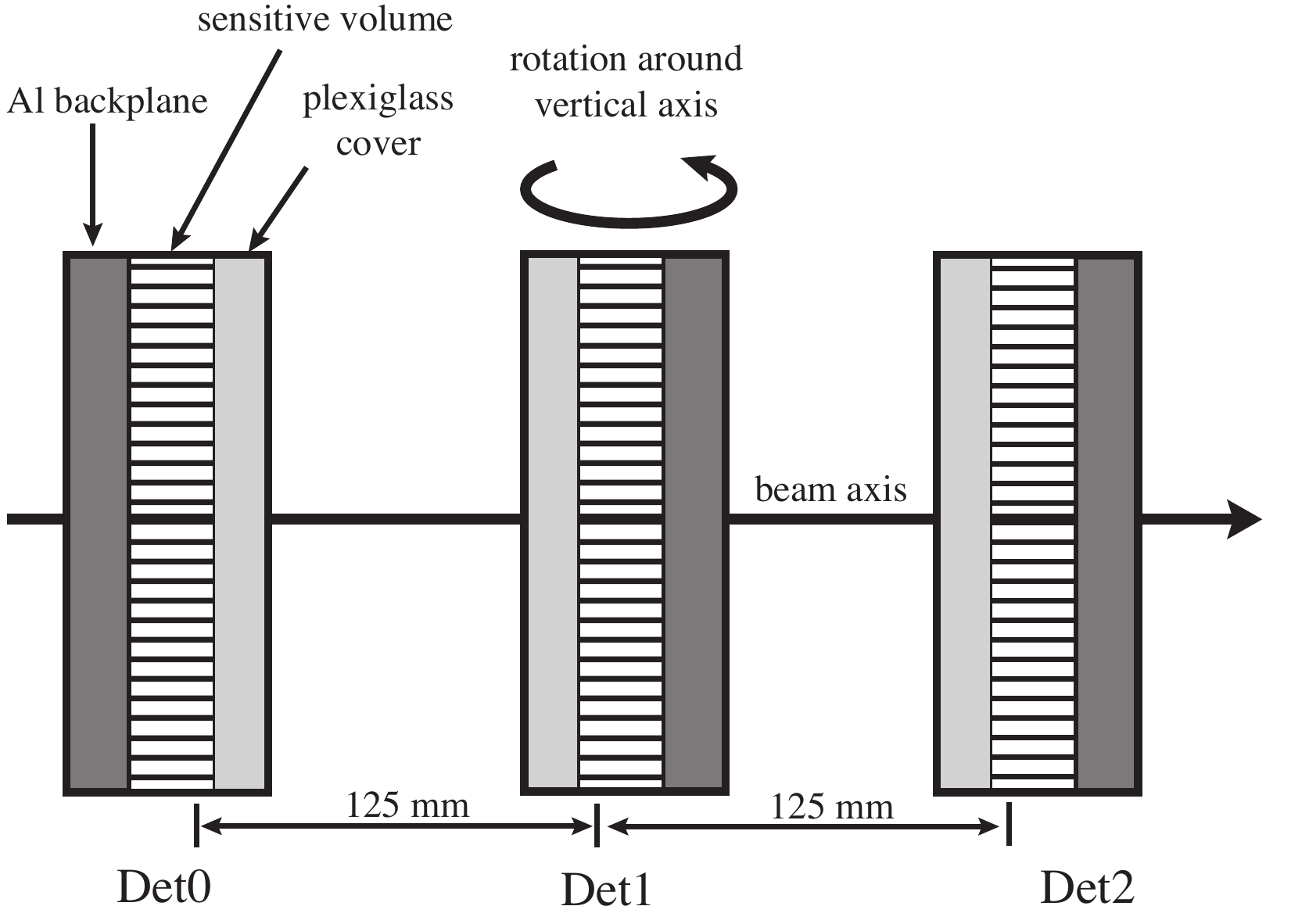}
\caption{The layout of the test beam installation. The three detectors were installed with 125 mm distance between the center of the drift gaps. To minimize the material within the tracking telescope, the back plane of the first detector faced the beam, while the back planes of the other two detectors faced downstream. The central detector could be rotated by up to 30$^\circ$ around the vertical axis.}
\label{fig:TestBeamLayout}
\end{figure}

Three such detectors were tested in the Meson Test Beam Facility (MTBF) at Fermi National Accelerator Laboratory together with MRPC prototypes \cite{T963MRPC} as experiment T963. The triple-GEM detectors were installed as a tracking telescope with 125 mm spacing between them, as illustrated in Figure \ref{fig:TestBeamLayout}. The detectors are labelled Det0, Det1 and Det2, counting from upstream. The horizontal coordinate, read out by the wide bottom strips of the readout board, is referred to as X coordinate and the vertical coordinate, read out by the narrow top strips, is referred to as the Y coordinate.  The middle and the last detector (Det1 and Det2) used strip readout boards with 25 $\mu$m vertical separation, while the first detector (Det0) used a board with 50 $\mu$m vertical separation. The middle detector could be rotated around the vertical axis to study the effect of track inclination on spatial resolution and efficiency. The effect of multiple scattering on tracks within the GEM telescope was minimized by installing the first detector with its bottom facing the beam, and the other two detectors with their entrance windows facing the beam. That way the amount of material between the active region of the first and the last detector was only around 4\% $X_0$, concentrated in the support plate and the readout board of the central detector. The beam trigger and timing information was derived from scintillators installed up- and downstream of the T963 experimental setup.

Data were taken under a variety of different beam conditions, with energies ranging from 4 GeV to 32 GeV for unseparated secondary beams, and with a 120 GeV primary proton beam. The detectors were running at total effective gains of around \mbox{3\,500}, as discussed above. Over a period of two weeks the detectors were operated in stable data taking mode without any problems. 

\section{Data Reconstruction and Analysis}

\begin{figure}
\centering
\includegraphics[width=0.49\textwidth]{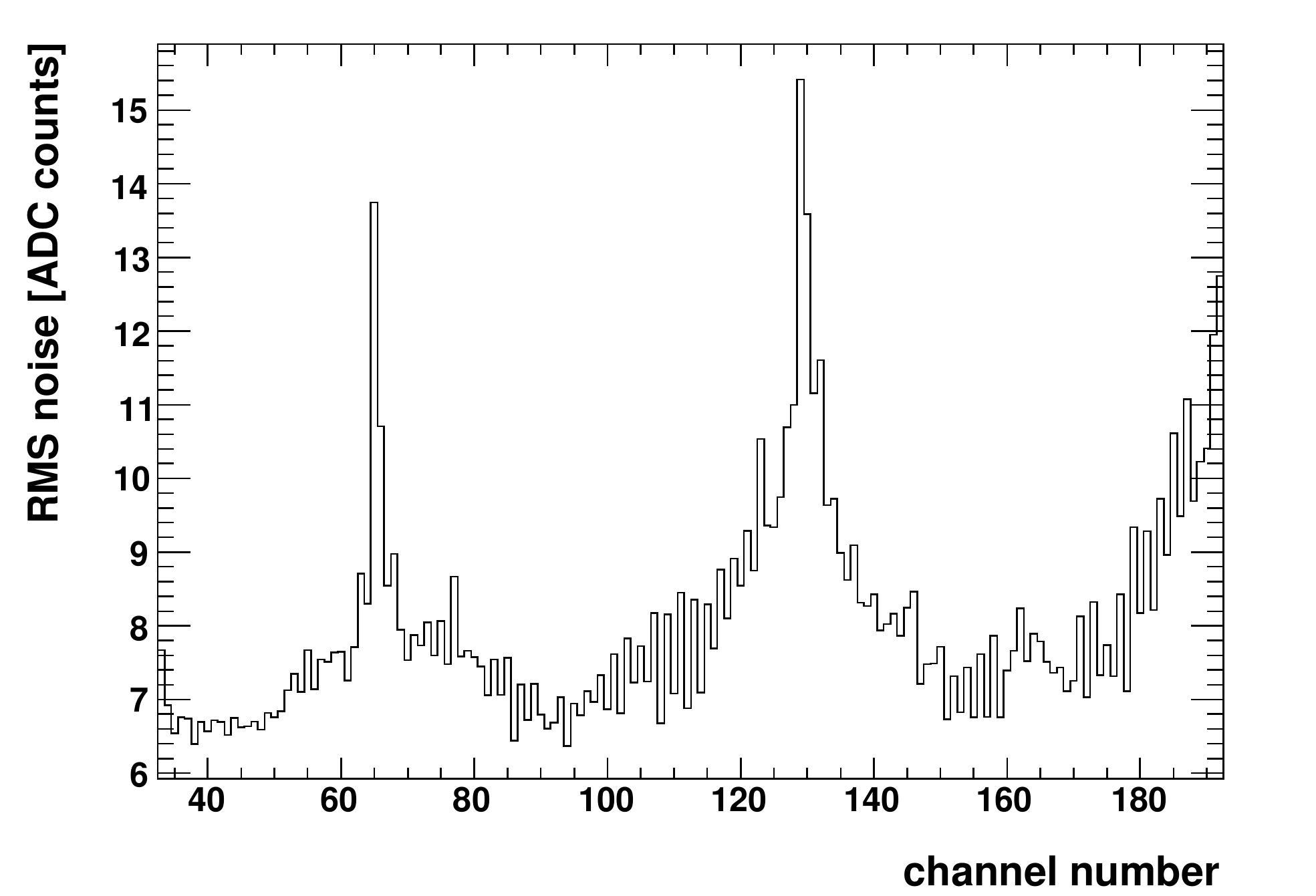}
\caption{Channel by channel noise distribution for the horizontal coordinate of the first detector. The first 32 channels are not connected, and not shown in the histogram. As only every other input channel is connected, the borders between chips are located at channel numbers 64, 65, 128, 129 and 192. These regions show elevated noise levels.}
\label{fig:Noise}
\end{figure}

The Fermilab main injector was delivering beam to the test beam area for a four second spill once per minute, with typically a few thousand particles per spill. For an intensity scan this was increased by about two orders of magnitude at the end of the test beam campaign, as discussed later. Data were taken in runs of around \mbox{10\,000} to \mbox{20\,000} events, at event rates of a few hundred to a thousand per spill. The data taking rate was limited by the readout time of the system. 
 
The data were stored in a not zero-suppressed format, with the raw amplitude of each electronics channel available for each event. It was observed that the amplitudes show signs of common mode noise (correlated shifts of the pedestal for all channels in one event), which was changing on a chip-by-chip basis. This common shift was determined via the median of the amplitudes of all channels in one chip and corrected for by shifting the baseline chip-by-chip in the data analysis. The observed common mode shifts were small in general, roughly following a Gaussian distribution with a width between $\sim$ 7 and $\sim$ 15 ADC counts, depending on the detector. These amplitudes correspond to  2\% to 6\% of a typical signal of a minimum ionizing particle. A fraction of about $10$\% to $12$\% of all events showed very large common mode shifts, which is attributed to pickup due to imperfect grounding and shielding. These events did not lead to significant reconstruction and analysis problems, but they occasionally exhibited saturation effects for the highest amplitude channels. 

Since the strip occupancy in the test beam was very low, on the order of 1\% to 3\%, the data events themselves were used to determine the channel pedestals on a run-by-run basis. 
The channel by channel pedestals and noise were determined from Gaussian fits to the distribution of the channel amplitudes after common mode noise subtraction. Figure \ref{fig:Noise} shows the distribution of the channel-by-channel noise for one detector projection. Typical noise levels were between 7 and 12 ADC counts, corresponding to an equivalent noise charge ranging from approximately $1\,250$ to $2\,200$ electrons. This relatively high noise level was due to the imperfect grounding and shielding of the electronics and readout system. In particular the regions at the chip borders show elevated noise levels.   

The pedestals were subtracted channel by channel for each event. 
The corrected channel amplitudes were then used to perform cluster finding, for each of the detector projections separately. The cluster finder was a simple peak finder which seeded the cluster at the strip with the highest amplitude. A minimum of $5\,\sigma_{noise}$ above pedestal, where $\sigma_{noise}$ is the standard deviation of the noise distribution of that particular channel, was required to form a seed. Adjacent strips were added into the cluster as long as their amplitude was more then $2\,\sigma_{noise}$ above pedestal. Cluster splitting was performed if there was a clear valley in the cluster. Single strip clusters are accepted. To suppress noise for the tracking studies discussed below, only clusters with two or more strips were accepted for the track defining detectors. For the measurements of the spatial resolution a minimum of two strips was imposed on all detectors. 

For all identified clusters, the space coordinate was determined from the center of gravity of the charge distribution within the cluster, and the total charge was determined by the integral of the signal over all strips in the cluster. The RMS cluster noise was determined using the $\sigma_{noise}$ values for each strip in the cluster, and defined as $\sigma_{cluster} = \sqrt{n^{-1} \sum \sigma^2_{noise}}$, where $n$ is the number of strips in the cluster, and the sum is taken over all strips in the cluster. Typical cluster noise values for hits on identified particle tracks are around 10 to 14 ADC counts. 

\begin{figure}
\centering
\includegraphics[width=0.49\textwidth]{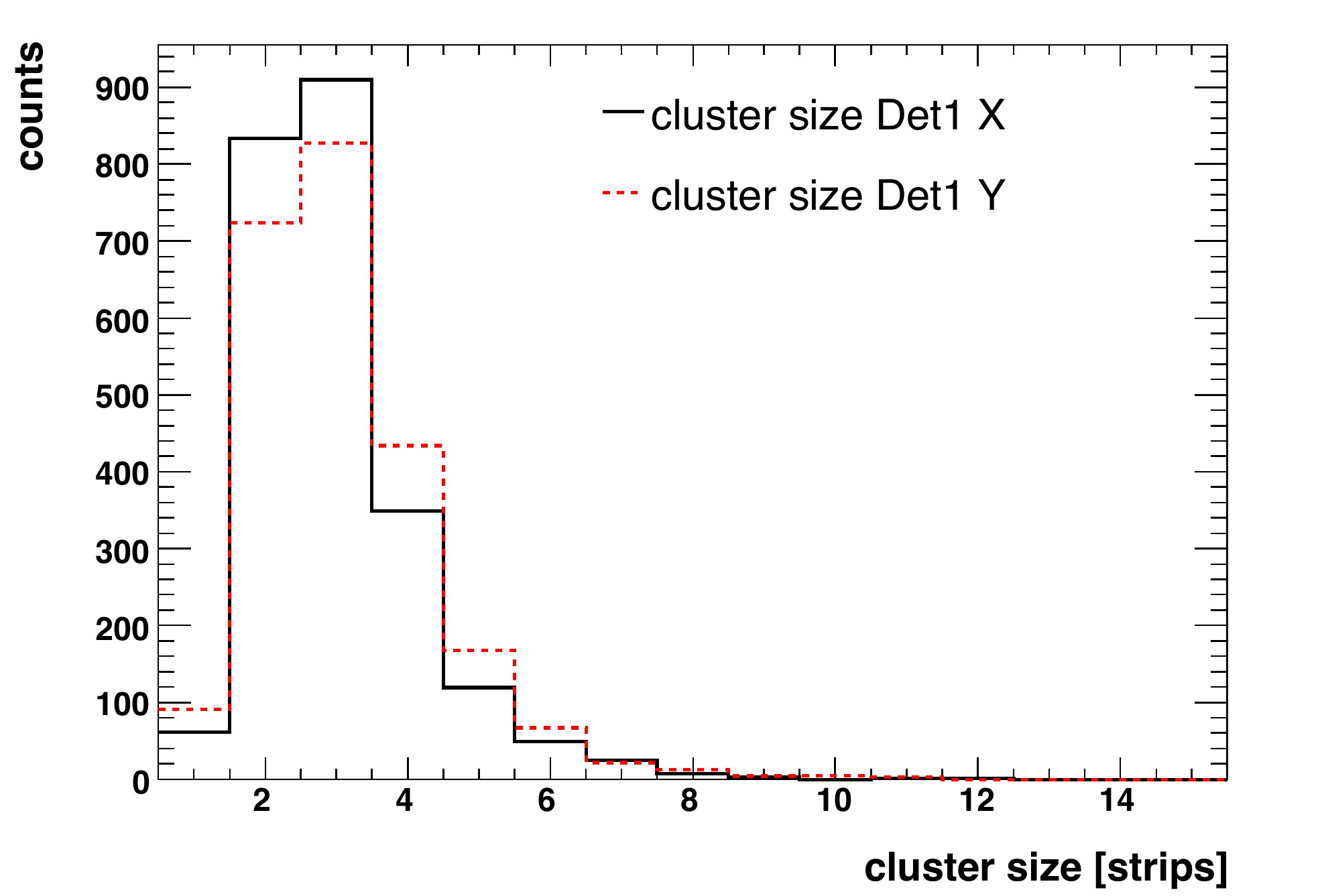}
\caption{The size of reconstructed clusters in strips (635 $\mu$m strip pitch) for both readout coordinates of the central detector.}
\label{fig:ClusterSize}
\end{figure}

Figure \ref{fig:ClusterSize} shows the cluster size in the middle detector in the tracking setup for clusters on tracks defined by the other two detectors. The most likely cluster size is between 2 and 3 strips on both coordinates, with a few percent of all found clusters being only one strip wide. This shows that a smaller strip pitch is desirable to obtain an optimal efficiency and the best possible spatial resolution.

\begin{figure}
\centering
\includegraphics[width=0.49\textwidth]{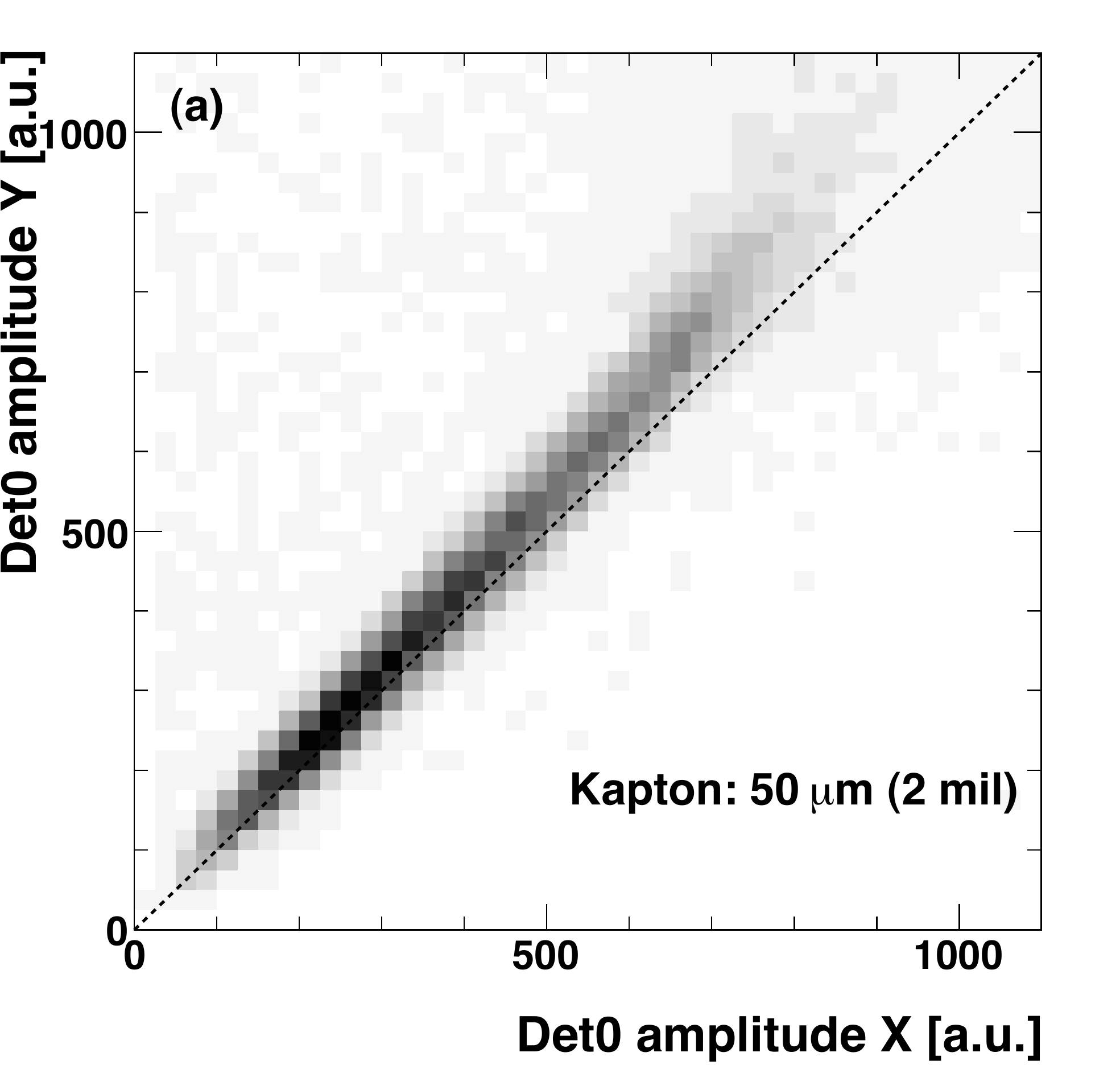}

\vspace*{10 mm}

\includegraphics[width=0.49\textwidth]{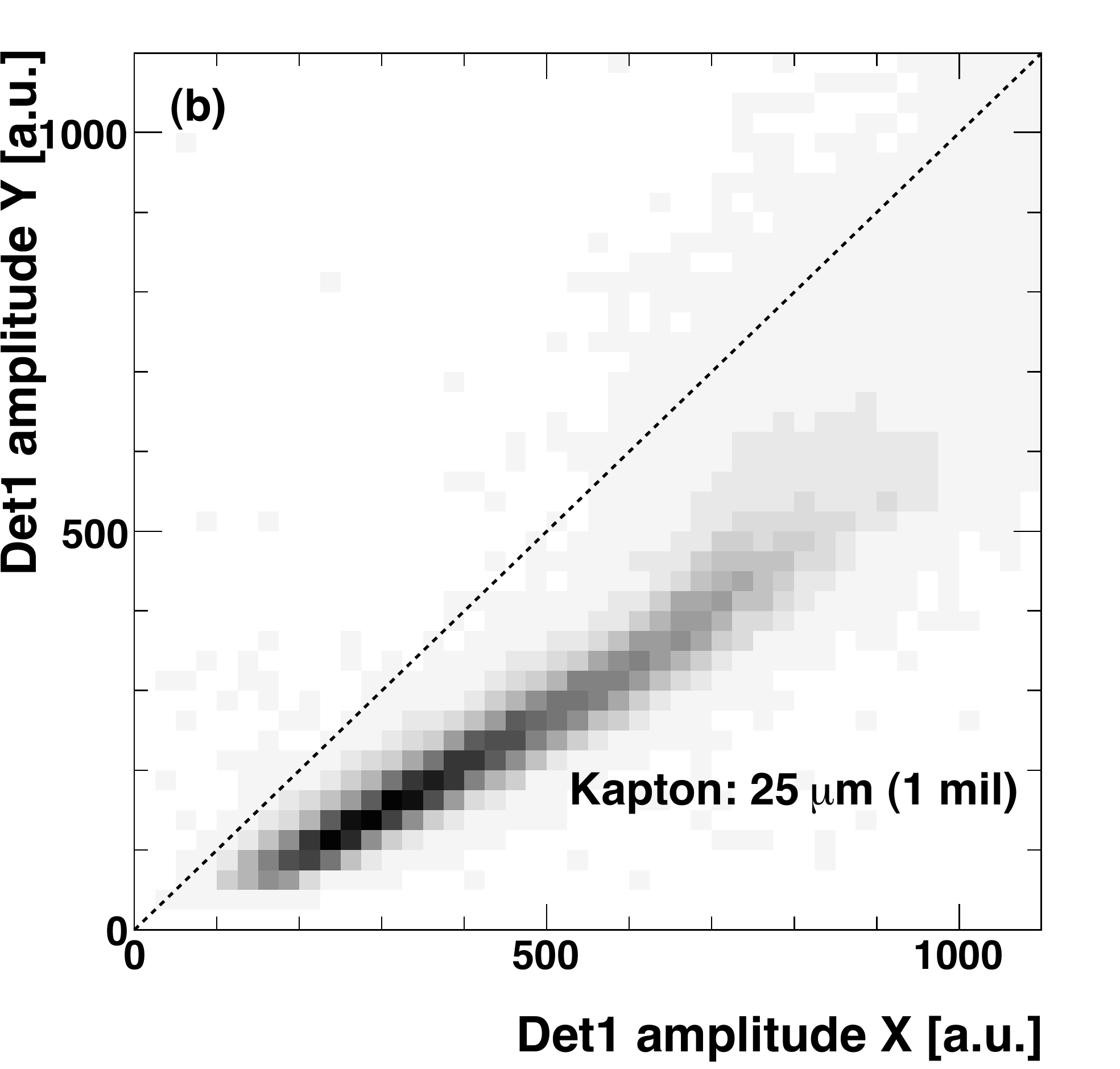}
\caption{Correlation of the reconstructed cluster charge between detector projections for the first detector (a) and the center detector (b) in the tracking setup. A charge sharing of 1:1 is indicated by the dashed line. The two different readout board designs show different charge sharing between coordinates. While the version with 50 $\mu$m vertical distance between the two strip planes achieves approximately equal charge sharing, the 25 $\mu$m version has a charge sharing of about 1:2 between the top and the bottom strips.}
\label{fig:AmplitudeCorrelations}
\end{figure}

Figure \ref{fig:AmplitudeCorrelations} shows the correlation of the reconstructed cluster charge between both readout coordinates for the two different versions of the readout board that were used in the beam test. The narrow correlations in both cases demonstrate a uniform sharing of the charge over the active area of the detectors. The sharing between coordinates is very different for the two versions of the readout board. While the version with 50 $\mu$m vertical distance between the two strip planes achieves approximately equal charge sharing, the 25 $\mu$m version has a charge sharing of about 1:2 between the top and the bottom strips. A Gaussian fit to the ratio of charge deposited on the horizontal coordinate to the charge on the vertical coordinate yields a mean of 1.09 and a sigma of 0.11 for Det0 (50 $\mu$m vertical distance) and a mean of 0.54 and a sigma of 0.087 for Det1 (25 $\mu$m vertical distance). 

\begin{figure}
\centering
\includegraphics[width=0.49\textwidth]{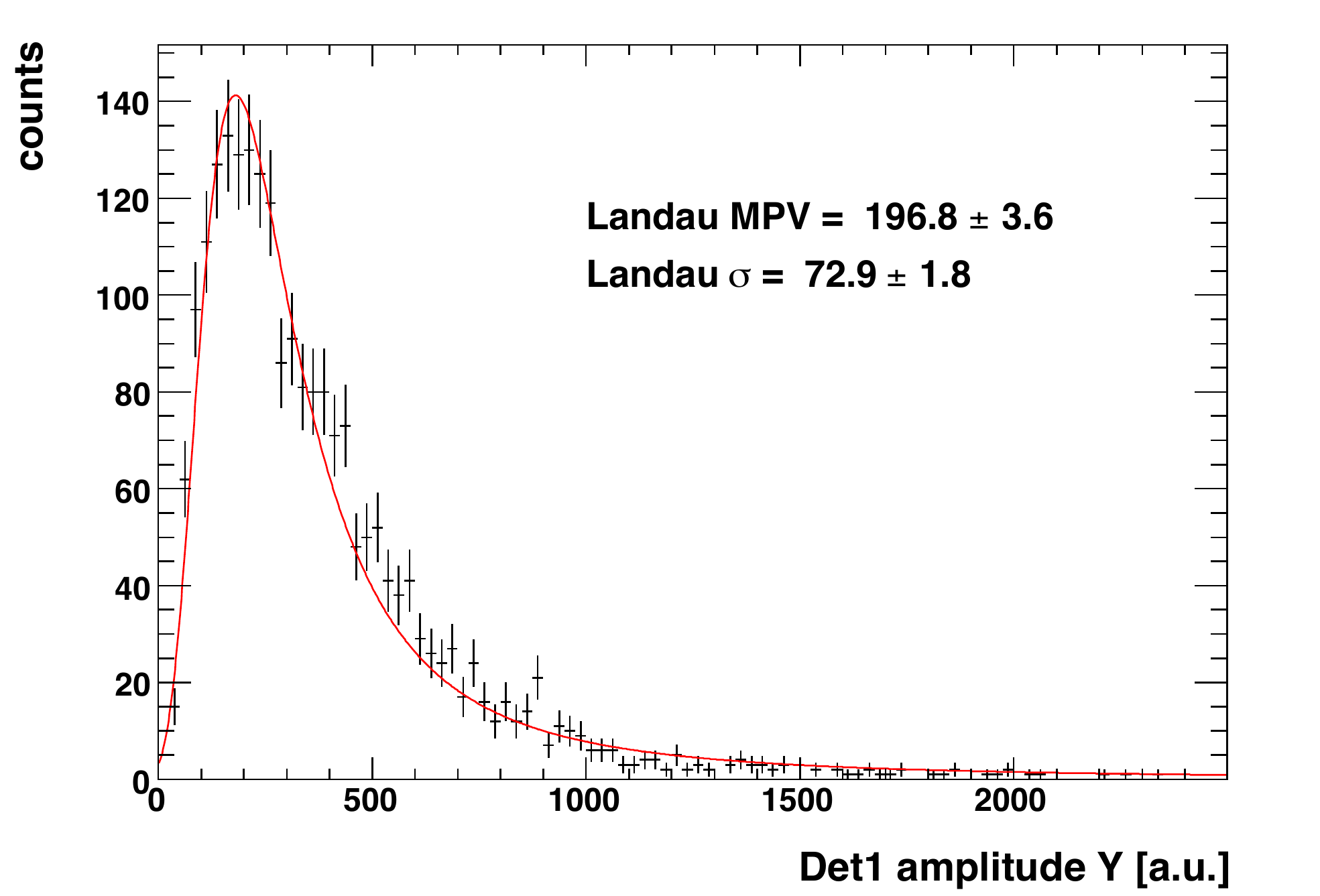}
\caption{Cluster charge distribution on one detector projection, together with a fit with a Landau function. The charge distribution follows the expected Landau shape for thin absorbers.}
\label{fig:AmplitudeLandau}
\end{figure}

Figure \ref{fig:AmplitudeLandau} shows the distribution of the reconstructed cluster charge on one projection of a detector. A Landau function is fitted to the distribution, showing the good agreement with the functional form expected for thin absorbers. The absence of a noise peak at low amplitudes is due to the strict requirement of at least $5\,\sigma_{noise}$ imposed on a strip to seed a cluster. 

\section{Tracking Performance}

The clusters found in the three detectors were used to study the tracking performance, namely efficiency, spatial resolution and the effect of track inclination, of the devices. Since the detectors were located in a field-free region, tracks were described as straight lines. Tracks were defined by the hits in two detectors to investigate the properties of the third detector not included in the track definition. To simplify the analysis procedure, only events with exactly one identified cluster on each projection of the two track-defining detectors were accepted. For studies of the spatial resolution, all three detectors in the telescope were required to have exactly one identified hit per projection. 

For these studies the data taken with a beam of 32 GeV and of 120 GeV were used to minimize the influence of multiple scattering. The uncertainty of the tracks in the central detector due to multiple scattering was about 11 $\mu$m for 32 GeV and 3 $\mu$m for 120 GeV. The size of the beam spot was strongly energy dependent. For the 32 GeV secondary beam, the beam spot had an approximately Gaussian shape with a sigma in the horizontal coordinate of $\sim$ 25.2 mm and a vertical sigma of $\sim$ 8.6 mm. In the horizontal plane, essentially the complete active area of the detectors was illuminated. In contrast, for the 120 GeV primary proton beam, the beam spot (even with maximum defocusing) was quite small, with a Gaussian sigma of $\sim$ 2.8 mm in horizontal and $\sim$ 3.9 mm in vertical direction. Due to the significantly smaller beam spot in the 120 GeV running, only a small region of the detectors was probed, reducing the influence from residual misalignments and from problematic detector areas at the borders of readout chips. Since the bulk of the data was taken in the 32 GeV configuration, the studies of the detector efficiency as a function of voltage and the studies of spatial resolution as a function of rotation angle of the central detector were performed with this energy. The spatial resolution without rotation of the detectors was measured with the 120 GeV proton beam. The 120 GeV beam also allowed the highest beam intensities, and was used to investigate rate effects in the detectors.

\subsection{Efficiency}

\begin{figure*}
\centering
\includegraphics[width=0.49\textwidth]{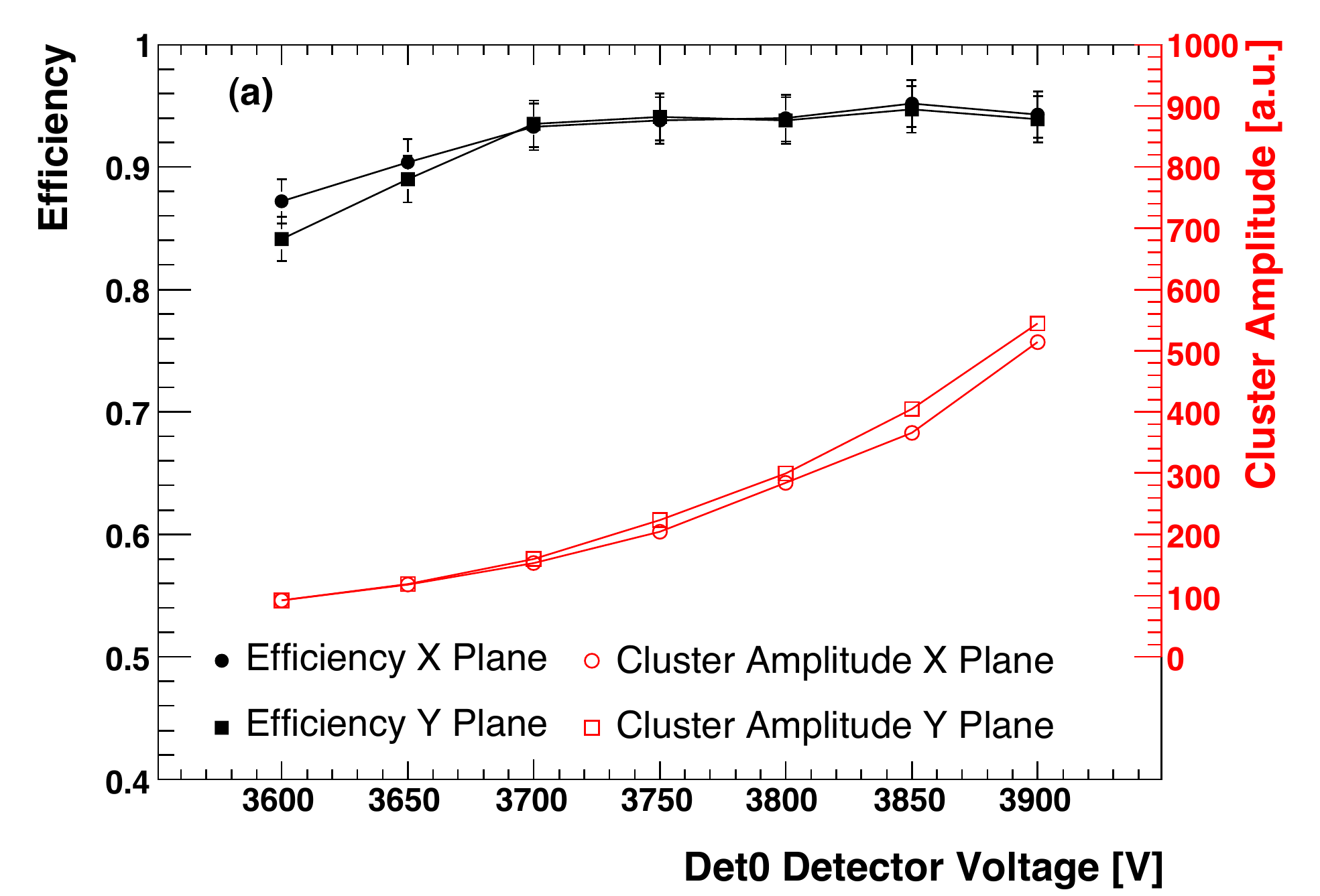}
\includegraphics[width=0.49\textwidth]{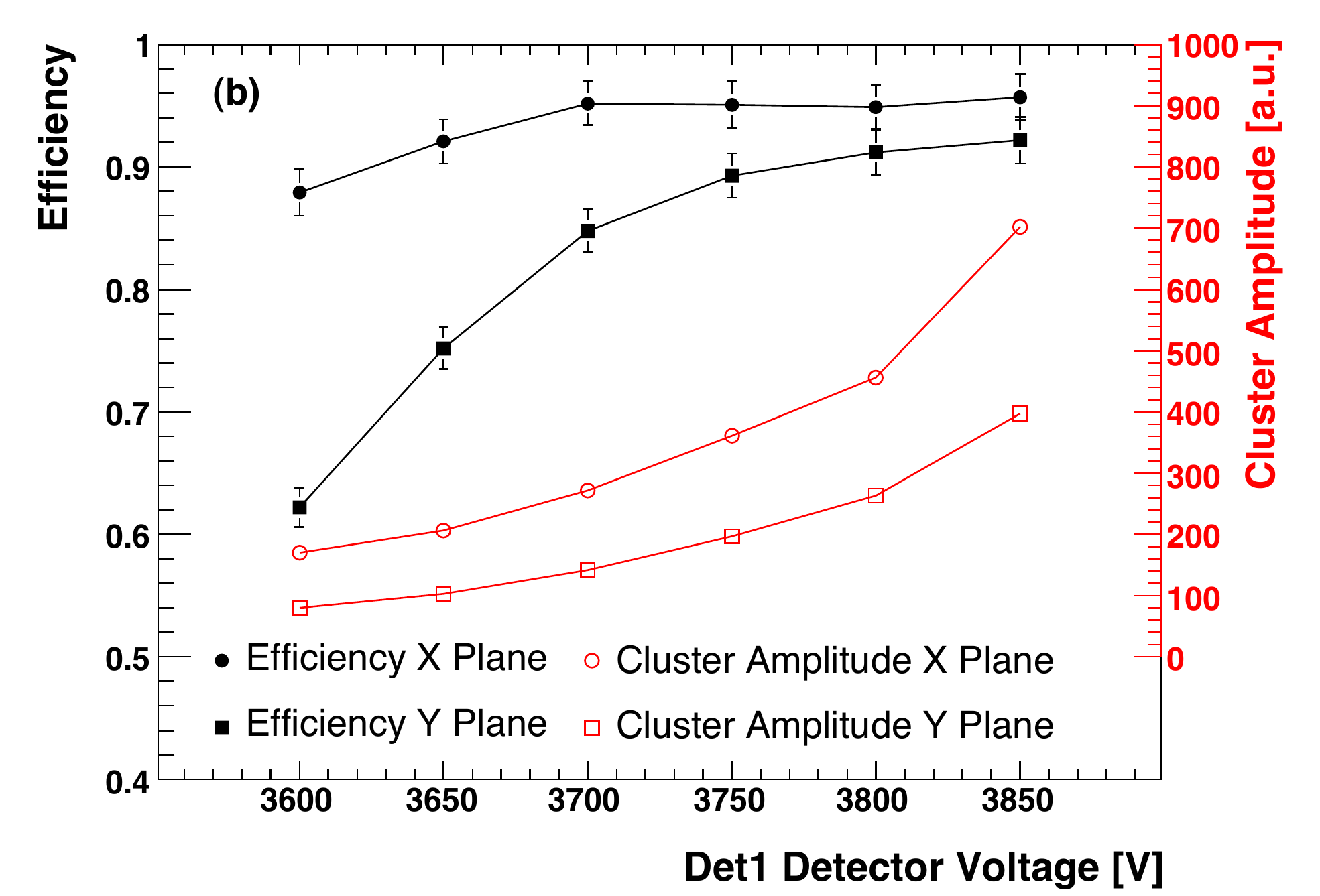}
\caption{Efficiency and cluster amplitude (most probable value for the cluster amplitude obtained by a Landau fit) as a function of detector voltage for the first (a) and second (b) detector in the tracking setup. The first detector used a readout board with 50 $\mu$m vertical separation between the strip layers, leading to a charge sharing of $\sim$ 1:1, while the second detector used 25 $\mu$m vertical separation, resulting in a charge sharing of $\sim$ 2:1 between the X and the Y plane.}
\label{fig:EfficiencyAmplitude}
\end{figure*}

Figure \ref{fig:EfficiencyAmplitude} shows the single-coordinate efficiency for the first (Det0) and middle (Det1) detector as a function of the applied voltage, obtained for the 32 GeV beam. A hit was counted as identified if it is found within 1 mm of the projected track position, using the standard cluster cuts as discussed above. It is apparent that both detectors reach an efficiency plateau at the higher voltages. Due to the inclusion of dead and noisy detector areas the plateau efficiency was reduced. For both projections on both detectors efficiencies between 95\% and 98\% were observed in restricted regions which limit the influence of dead and noisy areas, demonstrating that high efficiencies can be reached with tracking detectors using Tech-Etch GEM foils. Det0 used the readout board with approximately equal charge sharing, while Det1 had a 2:1 charge sharing between X and Y. Because of this, both coordinates of Det0 reach their plateau efficiency at the same voltage, while the vertical coordinate of Det1 reaches its plateau significantly later than the horizontal coordinate. 

Also shown in Figure \ref{fig:EfficiencyAmplitude} are the amplitudes of the clusters as a function of detector voltage, given by the most probable value of a Landau function fitted to the cluster charge distributions. Again, the effect of equal versus unequal charge sharing is clearly apparent in the differences between Det0 and Det1. An amplitude of $\sim$ 200 ADC counts is required to reach the efficiency plateau, corresponding to a cluster signal to noise ratio of $\sim$ 15 and a gain per coordinate of $\sim\ 1\,250$. Thus a total detector gain of $\sim\ 2\,500$ is necessary to reach the onset of the efficiency plateau for a detector with equal charge sharing, while a gain of $\sim\ 3\,750$ is necessary to reach the plateau onset for both coordinates in the case of 2:1 charge sharing. For stable running, both detectors shown here were operated at a gain of $\sim\ 3\,500$, reached at $3\,800$ V for Det0 and at $3\,750$ V for Det1. While Det0 was well into the efficiency plateau for both coordinates, Det1 was operated near the plateau onset for the vertical coordinate. The operation of Det1 at higher voltage was not practical however, since otherwise the horizontal coordinate would have run into saturation problems. This underlines the value of equal charge sharing for detectors with two-dimensional readout.

\subsection{Spatial Resolution}

\begin{figure}
\centering
\includegraphics[width=0.49\textwidth]{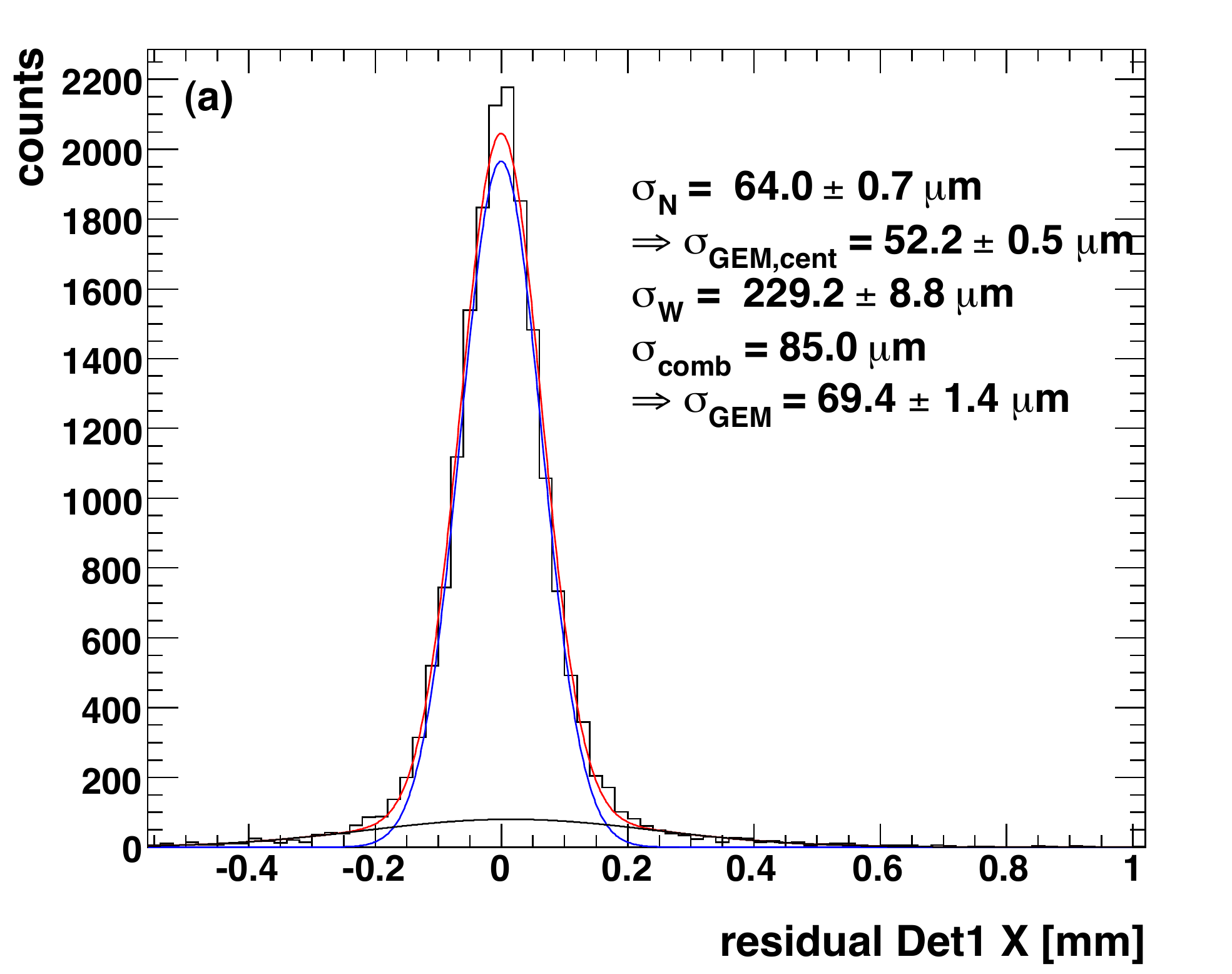}

\vspace*{10 mm}

\includegraphics[width=0.49\textwidth]{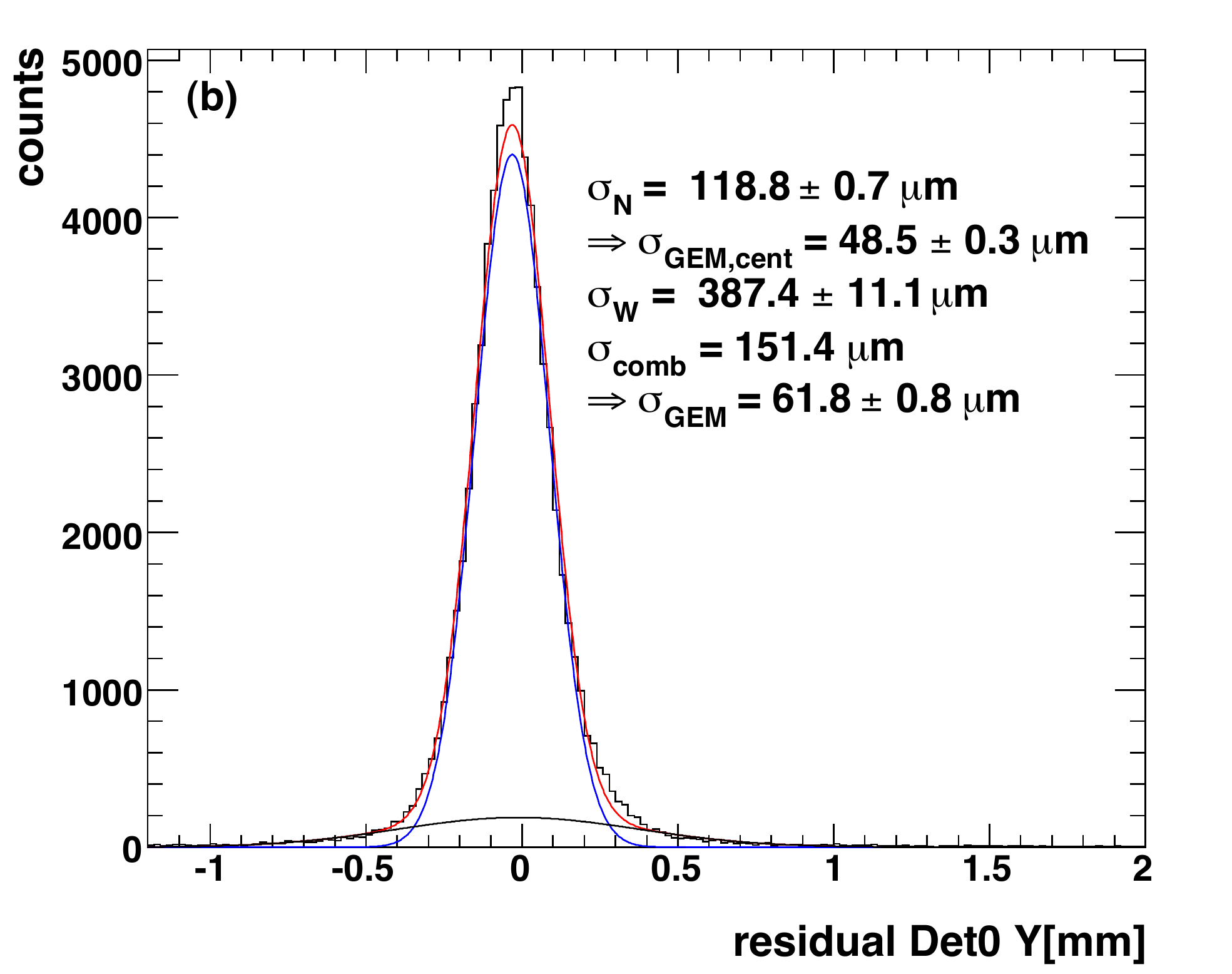}

\caption{(a) Residual distributions of  hits on the X projection of the central detector to tracks formed by the first and last detector in the tracking telescope and (b) residual distributions of  hits on the Y projection of the first detector to tracks extrapolated from the central and last detector in the tracking telescope. The resulting spatial resolution $\sigma_{GEM}$, extracted from a two-component Gaussian fit, is indicated. The residual distribution in (b) is considerably wider than in (a) due to the increased extrapolation uncertainty of the tracks.}
\label{fig:ResolutionX}
\end{figure}

From the distribution of the distance of the reconstructed hits in one detector to the tracks formed by the two other detectors the spatial resolution of the triple GEM test detectors was extracted. Figure \ref{fig:ResolutionX}a) shows this residual distribution for the X projection of the central detector and Figure \ref{fig:ResolutionX}b) shows the residual distribution for the Y projection of the first detector for data taken with the 120 GeV beam, where multiple scattering is negligible . The distributions are fitted with the sum of two Gaussians, a narrow one for the main central peak of the distribution, and one for the wider shoulders of the distribution. This combined function gives a very good description of the residual distribution for both detectors. The wider shoulders are probably due to one badly reconstructed hit in the track or in the detector under study, for example due to a dead or a noisy strip, or due to very low energy deposit. 

From the width of the residual distribution the spatial resolution of the detectors was determined by assuming that all three detectors have the same resolution, which is a reasonable assumption since the detectors are identical apart from the different versions of the readout board. The uncertainty of the track position, and thus the contribution of the track uncertainty to the measured residual, depends on how the track was formed. If the track was defined by hits in the first and last detector, the track uncertainty in the central detector is $\sigma_{track} = \sigma_{GEM} / \sqrt{2}$, where $\sigma_{GEM}$ is the spatial resolution of the GEM detectors. For the case where the track was defined by hits in the center and in the last detector, the track uncertainty in the first detector is $\sigma_{track} = \sigma_{GEM} \sqrt{5}$, considerably larger due to the extrapolation beyond the region covered by the track defining detectors.

In the first case for Det1 the spatial resolution is given by
\begin{equation}
\sigma_{GEM} = \sigma_{residual, 1}\,  \times\, \sqrt{\frac{2}{3}},
\end{equation}
while in the second case for Det0 it is given by
\begin{equation}
\sigma_{GEM} = \sigma_{residual, 0}\,  \times\, \frac{1}{\sqrt{6}},
\end{equation}

where $\sigma_{residual, 1}$ and $\sigma_{residual, 0}$ is the width of the residual distribution determined from the Gaussian fit in the central and in the first detector, respectively.

With the residual widths determined from the fits shown in Figure \ref{fig:ResolutionX} this results in a spatial resolution of $\sim$66 $\mu$m overall, using the weighted mean of the width of the narrow and the wide gaussian, and a resolution of $\sim$50 $\mu$m for the main central peak of the distribution. For studies with the 32 GeV beam and the correspondingly wider beam profile, resolutions around 65 $\mu$m for the main central peak and around 75 $\mu$m for the overall weighted resolution were obtained. This demonstrates that resolutions comparable to the ones obtained for the COMPASS triple GEM detectors \cite{Altunbas:2002ds,Ketzer:2004jk} are achievable with detectors based on Tech-Etch produced foils and laser-etched two dimensional readout. 

\subsection{Inclined Tracks}

In many applications the tracks do not have normal incidence to the plane of the detectors. This makes a study of the effect of track inclination on the detector performance crucial. In the test beam this was investigated by rotating the central detector by up to 30$^\circ$ around the vertical axis in the 32 GeV beam. As the primary interactions are statistically distributed along the particle track and follow Poissonian statistics in thin gaps, an inclination of the particle track leads to a smearing of the mean position of the charge, which in turn leads to a deterioration of the spatial resolution. This deterioration affects only the readout projection that is along the inclination of the track, the perpendicular projection is not affected. For the studies performed here, where the detector was rotated around the vertical axis, the resolution in the horizontal projection (X) is affected, while the resolution in the vertical projection (Y) should remain the same. 

\begin{figure}
\centering
\includegraphics[width=0.49\textwidth]{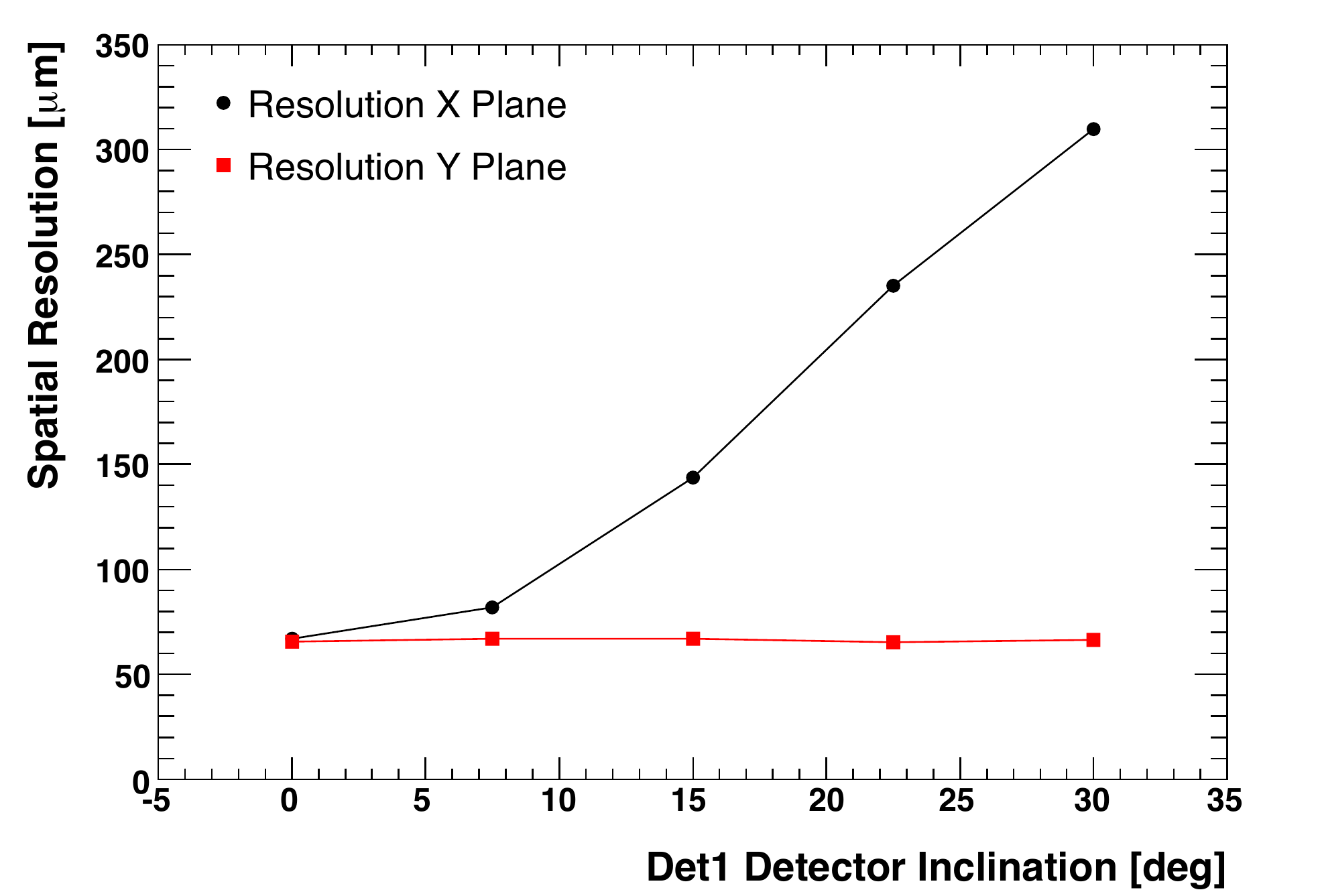}
\caption{Spatial resolution as a function of the angle of incidence of the particle tracks, for both the projection along the inclination of the tracks (X) and the projection perpendicular to the tracks (Y).}
\label{fig:ResolutionAngle}
\end{figure}

Figure \ref{fig:ResolutionAngle} shows the spatial resolution of both projections of the middle detector as a function of the angle of rotation around the vertical axis. As discussed above, the spatial resolution for the X projection deteriorates quickly with increasing angle while the spatial resolution in the Y projection remains unchanged. 

\subsection{Higher Rate Running}

At the end of the test beam period, the beam intensity was maximized to study the behavior of the GEM detectors when exposed to intense radiation. For this study the 120 GeV primary proton beam was used with the highest intensity permissible by the radiation safety limits in the test beam area. Data were taken with up to $\sim$ $1.6\cdot10^5$ protons within 4s spills corresponding to 40 kHz peak rate. With the size of the beam spot and the Gaussian profile discussed above, this corresponds to an intensity of a few kHz/mm$^2$ in the center of the beam spot. The time substructure of the spill was in bunches such that the peak intensities were probably considerably higher over very short periods, which is however not exactly known. The efficiency and the spatial resolution of the detectors measured at the highest intensities agree with the low intensity results within the measurement errors. The higher beam intensity thus led to no observable effects on the detector performance, and also did not induce any instabilities in the operation of the detectors.

\section{Conclusion}
Three test detectors using commercially produced GEM foils by Tech-Etch and a laser-etched two dimensional readout board produced by Compunetics have been tested extensively in a beam at Fermilab. The detectors showed stable performance during two weeks of beam operations. An efficiency in excess of 95\%  and a spatial resolution better than 70 $\mu$m was achieved, with a resolution of $\sim$ 50 $\mu$m for the majority of tracks. Two different versions of a laser-etched 2D orthogonal strip readout board were tested, giving important information on the charge sharing between readout coordinates depending on the board geometry. The detectors performed without problems at particle rates up to a few kHz/mm$^2$. Overall, the results of the beam test demonstrate that devices using commercially produced GEM foils from Tech-Etch and a laser-eched 2D readout achieve performance parameters comparable to those of triple-GEM detectors currently in operation in high energy physics experiments.

\section*{Acknowledgments}

The authors thank Fermilab for the allocation of beam time and Eric Ramberg and colleagues for generous support during the test beam activities. This work was supported in part by the U.S. Department of Energy, Division of Nuclear Physics, contract number DE-AC02-06CH11357 and by the U.S. Department of Energy, Division of Nuclear Physics, grant numbers DE-FG02-90ER40562 and DE-FG02-94ER40818.
The development of GEM foil production at Tech-Etch is supported by US-DOE SBIR grant DE-FG02-05ER84169.


\bibliographystyle{elsarticle-num.bst}

\bibliography{GEM}

\end{document}